\documentclass[conference]{IEEEtran}
\IEEEoverridecommandlockouts
\usepackage{cite}
\usepackage{amsmath,amssymb,amsfonts}
\usepackage{algorithmic}
\usepackage{graphicx}
\usepackage{textcomp}
\usepackage{xcolor}
\usepackage{amsmath}
\usepackage{float}
\usepackage{commath}
\usepackage{amssymb}
\usepackage{graphicx}
\usepackage{listings}
\usepackage{cite}
\usepackage{amssymb}
\graphicspath{ {figures/} }
\usepackage[colorinlistoftodos]{todonotes}
\def\BibTeX{{\rm B\kern-.05em{\sc i\kern-.025em b}\kern-.08em
    T\kern-.1667em\lower.7ex\hbox{E}\kern-.125emX}}
\begin{document}

\title{Demystifying the Role of zk-SNARKs in Zcash\\
}

\author{\IEEEauthorblockN{Aritra Banerjee}
\IEEEauthorblockA{
ADAPT Centre\\
\textit{Trinity College Dublin}\\
Dublin, Ireland \\
abanerje@tcd.ie
\and
\IEEEauthorblockN{Michael Clear}
\IEEEauthorblockA{
\textit{School of Comp Sci \& Stats}\\
\textit{Trinity College Dublin}\\
Dublin, Ireland \\
clearm@tcd.ie}
}
\and
\IEEEauthorblockN{Hitesh Tewari}
\IEEEauthorblockA{
\textit{School of Comp Sci \& Stats}\\
\textit{Trinity College Dublin}\\
Dublin, Ireland \\
htewari@tcd.ie}
}

\maketitle

\begin{abstract}
Zero-knowledge proofs have always provided a clear solution when it comes to conveying information from a prover to a verifier or vice versa without revealing essential information about the process. Advancements in zero-knowledge have helped develop proofs which are succinct and provide non-interactive arguments of knowledge along with maintaining the zero-knowledge criteria. zk-SNARKs (Zero knowledge Succinct Non-Interactive Argument of Knowledge) are one such method that outshines itself when it comes to advancement of zero-knowledge proofs. The underlying principle of the Zcash algorithm is such that it delivers a full-fledged ledger-based digital currency with strong privacy guarantees and the root of ensuring privacy lies fully on the construction of a proper zk-SNARK. In this paper we elaborate and construct a concrete zk-SNARK proof from scratch and explain its role in the Zcash algorithm.
\end{abstract}


\section{Introduction}
The first digital currency that attained worldwide recognition was Bitcoin, which was developed on the fact that it did not require a trusted third party. Contrary to the traditional e-cash schemes \cite{chaum1983blind} \cite{sander1999auditable}, Bitcoin used a distributed ledger called a \textit{blockchain} to store transactions. However Bitcoin is a ``pseudo-anonymous" scheme and does not provide for full anonymity of the participants. In 2013 Miers et al. proposed Zerocoin \cite{miers2013zerocoin} which extended the Bitcoin scheme by providing strong guarantees for anonymity. Zerocoin uses zero-knowledge proofs like many other e-cash protocols \cite{camenisch2005compact}. However, there were a few issues which existed with Zerocoin protocol: 1) The scheme only provided partial anonymity as it did not hide the amount or other important data in the transaction ledger but only hid the origin address; 2) The coins used had only fixed denominations; 3) The users could not pay each other directly using Zerocoins.

To address the above issues of both Bitcoin and Zerocoin, the Zerocash
algorithm was proposed by Ben-Sasson et al. \cite{sasson2014zerocash}. The concept of a decentralized anonymous payment scheme was introduced in Zcash which fundamentally provide the functionality and security guarantees and also ensures strong anonymity. The construction of the Zcash protocol is dependent on recent advances in zero-knowledge proofs, and specifically, zk-SNARKs \cite{groth2010short,lipmaa2012progression,ben2014succinct,groth2016size}.

\subsection{Overview of Zero-Knowledge Arguments of Knowledge}
An NP language is a set of strings $L$ such that if a string $\gamma$ belongs to the language $L$, then there exists a string $\alpha$, a witness that $\gamma$ belongs to $L$, that allows the membership of $\gamma \in L$ to be verified in polynomial time. We can thus define a polynomial-time computable relation $P$ such that $P(\gamma, \alpha) = 1$ if $\gamma \in L$ and $\alpha$ is a valid witness for $\gamma$. A non-interactive zero-knowledge argument of knowledge for an NP language $L$ allows a prover to convince a verifier that it has knowledge $\alpha$ that a given $\gamma$ belongs to $L$ without revealing anything about $\alpha$ to the verifier. A zk-SNARK\cite{bitansky2017hunting} is a non-interactive zero-knowledge argument of knowledge for NP that also achieves succinctness; that is, the argument (i.e. proof) is compact and computationally light to verify. However the security of these schemes are either based on non-falsifiable assumptions such as knowledge (extraction) assumptions or are instead proved secure in idealized models such as the generic group model.

The background idea used in Zcash while constructing the zk-SNARK is taken from the Pinocchio Protocol \cite{parno2013pinocchio}.

\begin{figure}[H]
  \centering
  \includegraphics[width=0.13\textwidth]{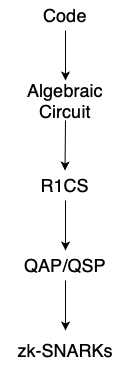}
  \caption{Conversion From Code to zk-SNARKs}
  \label{fig:steps}
\end{figure}

Figure \ref{fig:steps} illustrates the steps required to convert a piece of computer code into a zk-SNARK. In this paper we deal with each step in detail so as to give the reader a clear understanding of the inner workings of the zero knowledge protocol. Our work elaborates how code is converted to an algebraic circuit, which in turn gets converted to a Rank 1 Constraint System (R1CS) \cite{darouach1994full}. A R1CS is then converted to a Quadratic Arithmetic Program (QAP)/Quadratic Span Program (QSP) \cite{srivastava2010program}, and as a last step, a zk-SNARK proof\cite{DBLP:journals/corr/abs-1906-07221} is generated. Note that in our examples we use the field of real numbers for simplicity whereas in practice a ``large" finite field would be used. Further note that while we focus on Pinocchio in this paper, the zk-SNARK scheme used in Zcash at the time of writing is Groth's scheme \cite{groth2016size} from Eurocrypt 2016 (introduced in the sapling upgrade); the steps below from the code stage to QAP conversion remain unchanged.

\section{Steps to Create a zk-SNARK Proof}\label{zk-SNARK}
In a zero-knowledge proof, we can think of some program $P$ which takes as input a public input value $\gamma$ and witness $\alpha$, and outputs either 0 or 1. There are two main entities in this scenario: a \textit{prover} and a \textit{verifier}. Both prover and verifier know the public input $\gamma$. The prover has some special knowledge, also known as the witness $\alpha$, such that P($\gamma$, $\alpha$) = 1. Considering an example where $P$ (also mentioned as $f(x)$) is a sample cubic polynomial (derived from the Universal Circuit) as shown in Equation \eqref{eq1}:

\begin{equation}
    x^{3}+x+5 == 35
    \label{eq1}
\end{equation}

The above example and the idea for creating a simple zk-SNARK proof which we explain in this section (till subsection \ref{QAP}) was inspired by Vitalik Buterin's blog \cite{vitalik2016QAP}. The prover knows the proving polynomial (encrypted as the proving key), the witness and the public input. The verifier knows the public input, the zk-SNARK proof and the target polynomial (encrypted as verification key) to check whether prover has the correct polynomial (and hence the correct witness) or not. Basic algebra can be used to find that the value satisfying the equation is $3$. The main job of the prover is to prove that the function $f$ (defined in Section \ref{algebraic_circuit}) is executed correctly. In this scenario it is evident that the value of the witness $\alpha$ is $3$, the value of $P$ is the polynomial created in Equation \eqref{eq1}, the target polynomial $T$ is the polynomial form of Equation \eqref{eq9} and the public input $\gamma$ is $35$.

\subsection{High-Level Code}
A first step to any solution is to start with the high-level code representation of the problem in a programming language like C or Python. Translating Equation \eqref{eq1} into Python code we get the code snippet for the function definition of the polynomial equation.

\begin{lstlisting}[language=Python]
    def f(x):
        y = x**3
        return x + y + 5
\end{lstlisting}


\subsection{Algebraic Circuit}\label{algebraic_circuit}
The second step is the simple \textit{flattening} procedure wherein we convert the high-level code into a series of statements which contains expressions of the form: $x = y$ or maybe $x = y (op) z$. Here the value $op$ can be an addition, subtraction, multiplication or division operator, and $y$ and $z$ can be variables or numbers, or even further sub-expressions. Each of the statements after code flattening can be thought of as a \textit{gate} in the algebraic circuit. After the flattening procedure we obtain the following:

\begin{equation*}
    \begin{split}
        sym1 = x*x\\
        y = sym1*x\\
        sym2 = y+x\\
        out = sym2+5
    \end{split}
\end{equation*}

\begin{figure}[H]
  \centering
  \includegraphics[width=0.4\textwidth]{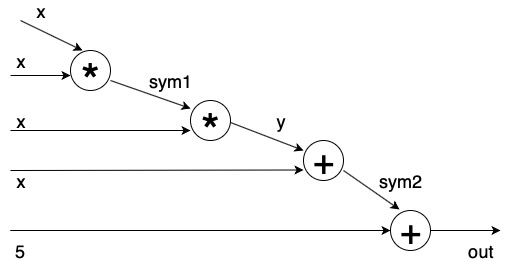}
  \caption{Algebraic Circuit Representation of the Flattened Code}
  \label{fig:gates}
\end{figure}

We observe that the flattened code and the original code are exactly the same. However we have now increased the number of lines and operators to form trivial algebraic gates. We can now represent the above flattened code as a system of gates as shown in Figure \ref{fig:gates}.

\subsection{Rank 1 Constraint System}
Next we convert the algebraic circuit into something called a Rank-1 constraint system (R1CS) \cite{darouach1994full}. A R1CS is a sequence of groups of three vectors $(v, w, k)$ where the solution to the R1CS is a vector $t$ such that it satisfies the equation:

\begin{equation}
    t . v * t . w - t . k = 0
    \label{eq2}
\end{equation}

\begin{figure}[H]
  \centering
  \includegraphics[width=0.4\textwidth]{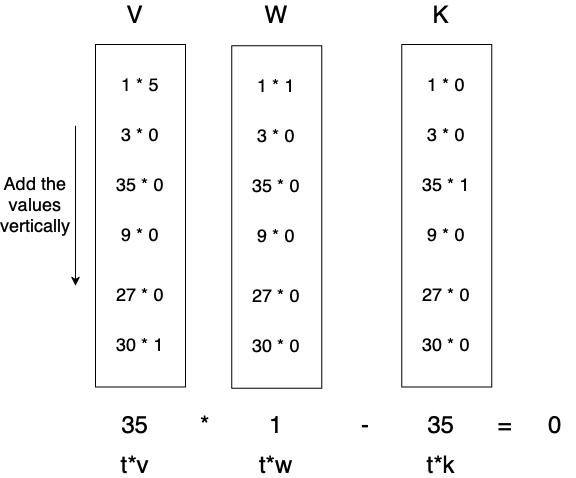}
  \caption{Satisfied R1CS System}
  \label{fig:r1cs}
\end{figure}

Where $(.)$ represents the dot product of the vectors. In simpler terms, if we join together $v$ and $t$, i.e. multiply the two values in the same positions and then take the sum of these products, then repeat the same to $w$ and $t$ and then $k$ and $t$, then the third result equals the product of the first two results. A satisfied R1CS is shown in Figure \ref{fig:r1cs}.

The value of the vector $t$ is $[1,3,35,9,27,30]$, which ensures a satisfied R1CS system. Instead of having just one constraint, we are going to have many constraints: one for each algebraic gate. There is a standard way of converting a  gate into a $(v, w, k)$ triple depending on what the operation is (+, -, * or /), and whether the arguments are variables or numbers. The length of each vector is equal to the total number of variables in the system. This includes a dummy variable \textit{one} at the first index representing the number 1, the input variables, a dummy variable $out$ representing the output, and then all of the intermediate variables ($sym1$ and $sym2$ as above). The vectors are generally going to be very sparse, only filling in the slots corresponding to the variables that are affected by some particular algebraic gate. The mapping we use is:

\begin{equation*}
    [one,x,out,sym1,y,sym2]
\end{equation*}

\noindent The solution vector consists of assignments for all of these variables, in the above given order. The triple $(v,w,k)$ for the first gate in Figure \ref{fig:gates} is given as:

\begin{equation*}
    \begin{split}
        v = [0, 1, 0, 0, 0, 0]\\
        w = [0, 1, 0, 0, 0, 0]\\
        k = [0, 0, 0, 1, 0, 0]\\
    \end{split}
\end{equation*}
\\

\noindent So, the solution $t = [1,3,0,9,0,0]$ indeed satisfies Equation \eqref{eq2}.
We assume that the solution vector has $3$ in the second position and $9$ in the fourth position, which satisfies the dot product $3*3 = 9$. If the value in the second position would have been $-3$; then also it would have passed having $9$ in the fourth position as $(-3)*(-3)$ is also $9$. In reality, $7$ in the second position and $49$ in the fourth position will also hold true because the purpose of the first check is just to verify consistency of the inputs and outputs in our case for the first multiplication gate only. Taking the second gate in Figure \ref{fig:gates} we check $y = sym1*x$. So the triple $(v,w,k)$ will be calculated in a similar fashion:

\begin{equation*}
    \begin{split}
        v = [0, 0, 0, 1, 0, 0]\\
        w = [0, 1, 0, 0, 0, 0]\\
        k = [0, 0, 0, 0, 1, 0]\\
    \end{split}
\end{equation*}
\\

\noindent The solution $t = [1,3,0,9,27,0]$ again satisfies Equation \eqref{eq2}.
Considering the third gate now:

\begin{equation*}
    \begin{split}
        v = [0, 1, 0, 0, 1, 0]\\
        w = [1, 0, 0, 0, 0, 0]\\
        k = [0, 0, 0, 0, 0, 1]\\
    \end{split}
\end{equation*}
\\

\noindent The solution $t = [1,3,0,9,27,30]$ satisfies Equation \eqref{eq2}.
In this case, the pattern is somewhat different. It’s multiplying the first element in the solution vector by the second element, then adding the fifth element to the results, and checking if the sum equals the sixth element. Because the first element in the solution vector is always one, this is just an addition check, checking that the output equals the sum of the two inputs.
Lastly for the fourth gate we have:

\begin{equation*}
    \begin{split}
        v = [5, 0, 0, 0, 0, 1]\\
        w = [1, 0, 0, 0, 0, 0]\\
        k = [0, 0, 1, 0, 0, 0]\\
    \end{split}
\end{equation*}
\\

The final solution becomes $t = [1,3,35,9,27,30]$ which satisfies Equation \eqref{eq2}. We are evaluating the check, $out = sym2 + 5$ in the fourth gate. The dot product check works by taking the sixth element in the solution vector, adding five times the first element (Note: the first element is 1, so this effectively means adding 5), and checking it against the third element, which is where we store the output variable.

\noindent Finally we have our R1CS with four constraints (i.e. for four logic gates). This is the witness which is the assignment of all variables, including input, output and internal variables: $t = [1,3,35,9,27,30]$. The final R1CS is:

\begin{equation}
    \begin{split}
        V
        [0, 1, 0, 0, 0, 0]
        [0, 0, 0, 1, 0, 0]
        [0, 1, 0, 0, 1, 0]
        [5, 0, 0, 0, 0, 1];
        \\
        W
        [0, 1, 0, 0, 0, 0]
        [0, 1, 0, 0, 0, 0]
        [1, 0, 0, 0, 0, 0]
        [1, 0, 0, 0, 0, 0];
        \\
        K
        [0, 0, 0, 1, 0, 0]
        [0, 0, 0, 0, 1, 0]
        [0, 0, 0, 0, 0, 1]
        [0, 0, 1, 0, 0, 0];
    \end{split}
    \label{eq3}
\end{equation}
\\

\subsection{Converting R1CS to Quadratic Arithmetic Programs}
In this step we take the R1CS and convert it into a Quadratic Arithmetic Program (QAP) \cite{srivastava2010program}, where we use similar concepts but use polynomials instead of dot products. In R1CS there are four groups of three vectors of length six. In a QAP, this gets converted to six groups of three degree-3 polynomials, where evaluating the polynomials at each $x$ coordinate represents one of the constraints, and we have four in our example as there are four gates. The main idea is that if we evaluate the polynomials at $x=1$, we get our first set of vectors. If we then evaluate the polynomials at $x=2$, we get our second set of vectors etc. This transformation can be made using Lagrange Interpolation \cite{nevai1984mean}. The problem that a Lagrange interpolation solves is that if you have a set of points (i.e. $\{x, y\}$ coordinate pairs), then doing a Lagrange interpolation on those points gives you a polynomial that passes through all of those points. This is done by decomposing the problem. For each $x$ coordinate, we create a polynomial that has the desired $y$ coordinate at that $x$ coordinate, and a $y$ coordinate of $0$ at all the other $x$ coordinates we are interested in. To get the final result we add all of the polynomials together. In mathematical terms, we explain the entire process for the $v$ vector. The goal is to construct $v_{1}(x)$ which is the first of the six polynomials in $V$. Initially we will be using the formula to create the $y$ values, i.e. the function points for Lagrange interpolation.

\begin{equation}
    v_{j}(i) = V[i][j]
    \label{eq4}
\end{equation}

\noindent In Equation \eqref{eq4}, the value $i$ is the vector index and $j$ is the position index in each vector for $V$ group in Equation \eqref{eq3}.The $x$ values are defined as $1\leq x \leq4$. Using Equation \eqref{eq3} we get the $y$ values as:

\begin{equation*}
    \begin{split}
        v_{1}(1) = 0\\
        v_{1}(2) = 0\\
        v_{1}(3) = 0\\
        v_{1}(4) = 5\\
    \end{split}
\end{equation*}

\noindent We already know the $x$ values as $1,2,3,4$. Using Lagrange interpolation formula \cite{nevai1984mean}:

\begin{equation}
    v_{1}(x) = 5* \frac{(x-1)}{4-1} \frac{(x-2)}{4-2} \frac{(x-3)}{4-3}
    \label{eq5}
\end{equation}

\noindent Solving Equation \eqref{eq5}:

\begin{equation*}
    v_{1}(x) = \frac{5}{6}x^{3} - 5x^{2} + \frac{55}{6}x - 5
\end{equation*}

\noindent Which can be also written in decimal form for ease of calculation as:

\begin{equation*}
    v_{1}(x) = 0.833x^{3} - 5x^{2} + 9.166x - 5
\end{equation*}

\noindent In similar way, we calculate $v_{2}(x),v_{3}(x),v_{4}(x),v_{5}(x),v_{6}(x)$ which sums up the entire $V$ group. After that we repeat the same for the $W$ group and $K$ group in Equation \eqref{eq3}. The six polynomials for the $V$ group looks like:

\begin{equation*}
    \begin{split}
        V\\
        [-5.0, 9.166, -5.0, 0.833]\\
        [8.0, -11.333, 5.0, -0.666]\\
        [0.0, 0.0, 0.0, 0.0]\\
        [-6.0, 9.5, -4.0, 0.5]\\
        [4.0, -7.0, 3.5, -0.5]\\
        [-1.0, 1.833, -1.0, 0.166]\\
    \end{split}
\end{equation*}
\\

Similarly we get six set of polynomials for $W$ group and the $K$ group respectively. This set of $(V,W,K)$ polynomials (plus a Z polynomial which is explained later) makes up the parameters for this particular QAP instance. Note that all of the work up until this point needs to be done only once for every function that one is trying to use zk-SNARKs to verify (such as in the case of a token transfer). Once the QAP parameters are generated, they can always be reused for that transaction.

\noindent If we try to evaluate the polynomials at $x=1$ we should get the set of $(v,w,k)$ vectors for the first logic gate.

\begin{equation*}
    \begin{split}
        \text{V results at }x = 1 \text{ } :
        [0,1,0,0,0,0]\\
        \text{W results at }x = 1 \text{ } :
        [0,1,0,0,0,0]\\
        \text{K results at }x = 1 \text{ } :
        [0,0,0,1,0,0]
    \end{split}
\end{equation*}

\noindent These values are exactly same as the first logic gate R1CS values. This property holds for all other values of $x$ as well.

\subsection{Need for Conversion}\label{QAP}
We need to convert R1CS to QAP so that we can check all the constraints simultaneously instead of checking the constraints in R1CS individually. We can now check all of the constraints at the same time by evaluating the dot product check on the polynomials. For ease of calculations let us consider $V(x) * W(x) - K(x)$ as $T$ (also known as the target polynomial). Here, $V(x)$ is:

\begin{equation}
    \begin{split}
        [t(1)]*v_{1}(x)\\ 
        +\\
        [t(2)]*v_{2}(x)\\ 
        .\\
        .\\
        [t(6)]*v_{6}(x) 
    \end{split}
    \label{eq6}
\end{equation}

\noindent Where $t$ is defined earlier as $[1,3,35,9,27,30]$. We repeat the same to get $W(x)$ and $K(x)$ in a similar fashion as designed in Equation \eqref{eq6}.
The goal is to find a polynomial $H$ such that the equation:

\begin{equation}
    T = H * Z(x)
\end{equation}

\noindent Such that the division $T/Z$ leaves no remainder. The value $Z$ is defined as the minimal simplest polynomial that is equal to zero at all points that correspond to logic gates. In this case we have four gates, so value of $Z$ becomes $(x-1)*(x-2)*(x-3)*(x-4)$. Performing polynomial multiplication we get:

\begin{equation}
    \begin{split}
        V(x) = [43.0, -73.333, 38.5, -5.166]\\
        W(x) = [-3.0, 10.333, -5.0, 0.666]\\
        K(x) = [-41.0, 71.666, -24.5, 2.833]\\
    \end{split}
    \label{eq7}
\end{equation}
\\

\noindent Using the Equation \eqref{eq7}, we get the value of $T = V(x)*W(x)-K(x)$ as:

\begin{equation}
\begin{split}
    T = [-88.0, 592.666, -1063.777, 805.833,\\ -294.777, 51.5, -3.444]
    \end{split}
    \label{eq9}
\end{equation}

\noindent Where $-3.444$ is the coefficient of $x^{6}$ term, $51.5$ is the coefficient $x^{5}$ term and so on. $Z$ is evaluated as:

\begin{equation}
    Z = [24,-50,35,-10,1]
    \label{eq10}
\end{equation}

\noindent Using \eqref{eq9} and \eqref{eq10}:

\begin{equation*}
    H = T/Z = [-3.666,17.055,-3.444]
\end{equation*}

\noindent Which means $H = -3.444x^{2}+17.055x-3.666$, and there is no remainder after the long division is performed.

\subsection{zk-SNARKs}
Having generated the zk-SNARK proof, we now address the verification part of the process. The prover has derived the polynomials $H$ and $Z$ as explained in Section \ref{QAP} using the polynomial $f(x)$. Hence, the zk-SNARK proof is an encrypted tuple $\pi$, where $\pi = [g^H,g^Z]$. The prover adds the value of $\pi$ onto the transaction ledger (explained later). The verifier obtains the proof $\pi$, the public input $\gamma$ from the transaction ledger, and the target polynomial $T$ (encrypted as the verification key $g^T$) from the public parameters string (\textit{ppar} - see Appendix \ref{DG}). The verifier knows only the public input $\gamma$, but not the polynomial $f(x)$.

Therefore, we use the concept of Homomorphic Hiding \cite{gentry2009fully} for the blind evaluation of a polynomial. Using Homomorphic Hiding and the ECIES encryption scheme \cite{koblitz2000state} (see Appendix \ref{DG}), the verifier calculates the values of $E(1),E(\gamma),E(\gamma^{2}),E(\gamma^{3}),E(\gamma^{4}),E(\gamma^{5}),E(\gamma^{6})$. The verifier then computes the values of $E(H(\gamma))*E(Z(\gamma))$ (verifier received $[g^H,g^Z]$ from the proof $\pi$) along with the value of $E(T(\gamma))$. This can be easily done because the encryption $E$ supports linear combinations, i.e. the values $H(\gamma)$, $Z(\gamma)$ and $T(\gamma)$ are a linear combination of $1,\gamma,\gamma^{2},\gamma^{3},\gamma^{4},\gamma^{5},\gamma^{6}$.

If the value of $E(H(\gamma))*E(Z(\gamma))$ (received from the proof $\pi$) matches the value of $E(T(\gamma))$ (target polynomial received as the verification key $g^T$), then the verifier is convinced that the prover knows the right polynomial, and hence the right witness. Given for a field $\mathbb{F}$, a zk-SNARK for $\mathbb{F}$-arithmetic circuit satisfiability is a triple of polynomial time algorithms (\textit{KeyGen}, \textit{Prove}, \textit{Verify}). In Zcash, zK-SNARKs based on QAPs are the main basis of construction; as this provides a linear time \textit{KeyGen} function, quasi-linear time \textit{Prove} function and linear time \textit{Verify} function.

\section{Role of zk-SNARKs in Zcash}
\begin{figure}[H]
  \centering
  \includegraphics[width=0.45\textwidth]{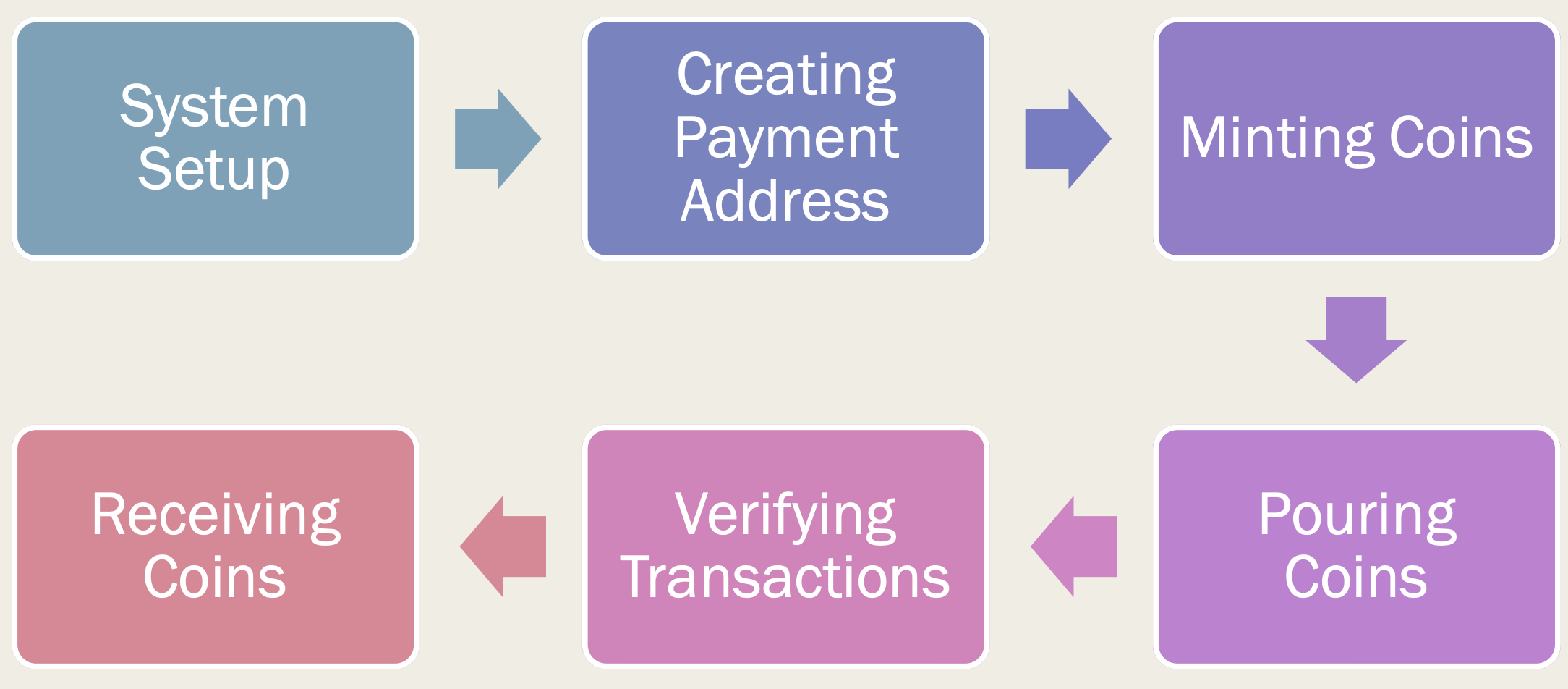}
  \caption{Steps Involved in a Zcash Transaction}
  \label{fig:zcash}
\end{figure}

Finally, we elaborate on how the entire zk-SNARK mechanism (as explained in Section \ref{zk-SNARK}) can work in a Zcash transaction \cite{parno2013pinocchio}. Figure \ref{fig:zcash} illustrates the six steps that is involved in any Zcash transaction.

The zk-SNARK proofs are usually generated in the \textbf{Pouring Coins} phase \cite{sasson2014zerocash}. Initially, we have a security parameter $\lambda$, which can be viewed as the number of bits of security and a Universal Circuit $C$ (usually a million gated circuit in Zcash). During the \textbf{System Setup} phase, using the value of $C$ and $\lambda$, the \textit{KeyGen} function \cite{boneh1997efficient,frankel1998robust} samples the proving key (encrypted proving polynomial) and the verification key (encrypted target polynomial) and stores it as the output. This is done using the RSA \cite{rivest1978method} protocol by the \textit{Trusted Third Party} \cite{blum1983exchange,demillo1983protocols}. These are stored in the \textit{ppar} string which can then be easily accessed by both the \textit{prover} and the \textit{verifier} (since the \textit{ppar} string is stored on the Blockchain). \textbf{Note}: Since Zcash employs zk-SNARKs for its knowledge construction, it implies that Zcash is essentially a \textit{publicly verifiable} mechanism \cite{sasson2014zerocash,parno2013pinocchio} and hence anyone can be the \textit{verifier}.

In the \textbf{Creating Payment Address} phase, the public encryption scheme (ECIES encryption) is used to create a public key and a secret key for a single user. These keys are in turn as used in conjunction with a pseudo-random function 
\cite{luby1988construct} (PRF with a seed value) to generate a public/private key address pair (see Appendix \ref{DG}). 

Next comes the \textbf{Minting Coins} phase where, as the name suggests, new coins are minted from a user's known coin addresses. The inputs are the \textit{ppar} string, the max value of a coin and the public key address (output of the Creating Payment Address phase) where the newly minted coin will be stored. This phase provides an output tuple which stores the coin value, the public address of the coin and the \textit{coin commitment} schemes \cite{brassard1993quantum}, created using randomly sampled trapdoors \cite{boyle2014functional} (based on a PRF using the coin serial number).

Following the \textbf{Minting Coins} phase we get to the \textbf{Pouring Coins} phase which is basically used to pour the values of old coins into new coins, thus ensuring that the old coins have been used. This phase takes as input \textit{n} (in Zcash the value of \textit{n} = $2$) distinct coins and their \textit{n} distinct secret key addresses. It also takes as input the \textit{ppar} string, the \textit{Merkle tree root} \cite{szydlo2004merkle, jakobsson2003fractal} (to the ensure that \textit{n} coins have been minted), and also \textit{n} authentication paths for the \textit{n} coin commitments ($cm_{1}(c^{old}_{1}),cm_{2}(c^{old}_{2})...,cm_{n}(c^{old}_{n})$) \cite{brassard1993quantum}. The new input values of the new \textit{n} coins and their public key addresses are additionally taken as inputs. Internally during the \textbf{Pouring Coins} phase, the \textit{prover} executes the \textit{Prove} function and generates the zk-SNARK proof based on an NP statement using the proving key ($f(x)$ in our example), the witness ($\alpha$ in our example) and the public input ($\gamma$ in our example), which are all part of the NP statement (see Appendix \ref{DG}). The zk-SNARK proof is added onto the \textit{POUR} transaction ledger (Note: There are two transaction ledgers namely \textit{MINT} and \textit{POUR}). This transaction ledger and the new \textit{n} coins are stored as the output.

Subsequently in the \textbf{Verifying Transactions} phase, the \textit{verifier} either verifies the \textit{Minting coin} transaction ledger or the \textit{POUR} transaction ledger. This phase takes as input the \textit{ppar} string, either the \textit{MINT} or the \textit{POUR} transaction ledger and a current ledger. If the input is the \textit{MINT} transaction ledger, then the \textit{Verifier} just checks whether the committed coins have the same value as claimed by the \textit{Prover}. If the input is the \textit{POUR} transaction ledger, the \textit{Verifier} executes the \textit{Verify} function using the zk-SNARK proof ($\pi$), the public input ($\gamma$) as stored inside the \textit{POUR} transaction ledger and the verification key ($g^{T}$) (part of the \textit{ppar} string). The \textit{Verify} function returns a Boolean value depending on whether the \textit{verifier} is convinced that \textit{prover's} assertion is genuine. If the \textit{verifier} is convinced of the zk-SNARK proof and that the digital signature (see Appendix \ref{DG})
\cite{rivest1978method,johnson2001elliptic} is valid, the \textbf{Verifying Transactions} phase returns \emph{TRUE}. \textbf{Note}: This phase can also return a \emph{FALSE} value if the old serial number of coins are present on the current ledger, or the Merkle tree root \cite{szydlo2004merkle} is not present on the current ledger, as this would mean a user is trying to re-use spent coins.

Finally, we have the \textbf{Receiving Coins} phase. In this phase the \textit{receiver} is a user with an address pair (public \& private key) that wants to receive payments to be sent to the public key address. To do so one first scans the current ledger. For every payment to the public key address, the receiver receives coins whose serial number do not appear on the current ledger (i.e, this ensures that receiver receives only unspent coins). To spend the \textit{received} coins one needs to use the \textit{POUR} algorithm, thus creating the whole need to form a zk-SNARK proof, and subsequently the need to \textit{verify} the proof to ensure that a user is not re-using old (already spent) coins repeatedly (see Appendix \ref{BO} for a detailed diagram).

\section{Conclusion}
The method explained in this paper is the fundamental basis used in the construction of zk-SNARK proofs in the Zcash \cite{sasson2014zerocash} paper. It elaborates on the fact that the verifier is thinking of a \textit{right} polynomial, and wants to check whether the prover knows it. The prover generates a zk-SNARK proof from the proving key supplied by the public parameters string, along with the witness and the public input. The verifier has the \textit{target} polynomial (from public parameters string as verification key) with which to verify the proof. Therefore, if the prover does not have the correct \textit{proving} polynomial they will add a wrong proof to the public parameter string. So, it is very easy for the verifier to verify whether the prover knows the right proving polynomial or not, without the verifier actually knowing the polynomial itself.\\
\\
\section*{Acknowledgements} This publication has emanated from research conducted with the financial support of Science Foundation Ireland under Grant Number 13/RC/2106 (ADAPT) and 17/SP/5447. This work was also supported, in part, by Science Foundation Ireland grant 13/RC/2094 (Lero).

\bibliographystyle{./bibliography/IEEEtran}
\bibliography{./bibliography/IEEEabrv,./bibliography/AINS2020}
\appendix
\begin{appendices}
\section{Definitions and Glossary}\label{DG}
\begin{itemize}
    \item \textbf{Elliptic Curve Integrated Encryption Scheme (ECIES)}: During the encryption phase in this scheme, the public key of the receiver and the original message (to be encrypted) are taken as inputs. It results in an output string that consists of the public key of the sender, the encrypted message and a MAC (to check the authenticity of the sender) tag \cite{bellare2000security}. The decryption phase takes as input the private key of the receiver and the output string of the encryption phase. The resulting output after decryption is the original message.\newline
    
    \item \textbf{Homomorphic Hiding}: We have defined $E(x)$ as $g^{x}$; where $g$ is a generator of a group with a hard discrete log problem. Homomorphic Hiding supports addition in the sense that $E(x+y)$ can be calculated from $E(x)$ and $E(y)$; which means for the linear combinations of polynomials we obtain:
    \begin{equation*}
    \begin{split}
        E(ax+by) = g^{ax+by} = g^{ax}g^{by}\\
        = (g^{x})^{a}(g^{y})^{b} = E(x)^{a}E(y)^{b}
    \end{split}
    \end{equation*}\newline
    
    \item \textbf{Merkle Tree} : A Merkle tree is a binary tree where every leaf node is a block of data (a cryptographic hash), and every internal node (including the root) acts as a marker which contains the address (also a cryptographic hash) of its children.\newline
    
     \item \textbf{NP Statement}: For Zcash, the NP statement states that:\\
    With the given values of the Merkle tree root $rt$, the new coin commitments ($cm_{1}^{new},cm_{2}^{new}$) and the old coin's serial number ($sn^{old}$). The user \textbf{u} generates a zk-SNARK proof for the \textit{POUR} phase such that he can prove that he knows the values of the old ($c^{old}$) and new coins ($c_{1}^{new},c_{2}^{new}$), along with the secret address key of the old coin ($a_{sk}^{old}$). \\
    The above NP statement is mapped as the polynomial $f(x)$ in our paper. $rt, cm_{1}^{new},cm_{2}^{new}, sn^{old}$ are the \textit{public inputs} which are mapped as $\gamma$, and $a_{sk}^{old}$ is the \textit{witness} which is mapped as $\alpha$.\newline
    
    \item \textbf{ppar} : The public parameters string which consists of the proving and verification keys needed to prove and verify the \textit{POUR} transaction.\newline
    
    \item \textbf{Pseudo-random Function} : A PRF or pseudo-random function is an efficient deterministic keyed-function that given a key maps an arbitrary string to a pseudo-random value.\newline
    
    \item \textbf{Signature Scheme}: The ECDSA digital signature scheme is used.
\end{itemize}

\onecolumn \section{Brief Overview of steps in Zcash}\label{BO}
\begin{figure}[H]
  \centering
  \onecolumn \includegraphics[width=0.9\textwidth]{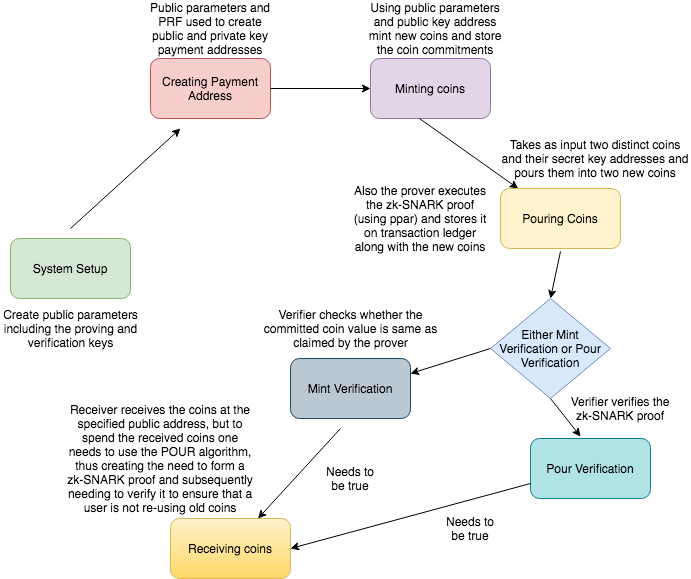}
  \label{fig:zcashdet}
\end{figure}
\end{appendices}

\end{document}